\documentclass{article}

\usepackage{arxiv}

\usepackage[utf8]{inputenc} 
\usepackage[T1]{fontenc}    
\usepackage{hyperref}       
\usepackage{url}            
\usepackage{booktabs}       
\usepackage{amsfonts}       
\usepackage{nicefrac}       
\usepackage{microtype}      
\usepackage{lipsum}		
\usepackage{graphicx}
\usepackage[square , sort, numbers]{natbib}
\usepackage{doi}
\usepackage[nolist]{acronym}
\usepackage{threeparttable} 
\usepackage{tabularx} 
\usepackage{tabularx}
\usepackage{multirow}
\usepackage{threeparttable}

\title{Impact of Benign Connectivity Variations on Intrusion Detection for Encrypted OPC~UA Traffic in Industrial Private~5G Networks}

\date{July 10, 2026}	

\author{ \href{https://orcid.org/0000-0003-1716-5867}{\includegraphics[scale=0.06]{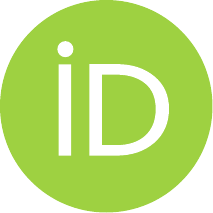}\hspace{1mm}Song Son Ha}\\
	Electrical Measurement Engineering\\
	Helmut-Schmidt-University\\
	Hamburg, Germany\\
	\texttt{song.ha@hsu-hh.de} \\
	\And
	\href{https://orcid.org/0009-0003-4437-010X}{\includegraphics[scale=0.06]{orcid.pdf}\hspace{1mm}Florian Foerster}\\
	Institute for Innovative Safety and Security\\
	Technical University of Applied Sciences Augsburg\\
	Augsburg, Germany\\
	\texttt{florian.foerster@tha.de} \\
	\And
	\href{https://orcid.org/0000-0002-5390-3946}{\includegraphics[scale=0.06]{orcid.pdf}\hspace{1mm}Henry Beuster}\\
	Electrical Measurement Engineering\\
	Helmut-Schmidt-University\\
	Hamburg, Germany\\
	\texttt{henry.beuster@hsu-hh.de} \\
	\And
	{\hspace{1mm}Tim Kittel}\\
	ipoque GmbH\\
	A Rohde \& Schwarz company\\
	Leipzig, Germany\\
	\texttt{tim.kittel@rohde-schwarz.com} \\
	\And
	\href{https://orcid.org/0000-0003-2310-5895}{\includegraphics[scale=0.06]{orcid.pdf}\hspace{1mm}Dominik Merli}\\
	Institute for Innovative Safety and Security\\
	Technical University of Applied Sciences Augsburg\\
	Augsburg, Germany\\
	\texttt{dominik.merli@tha.de} \\
	\And
	{\hspace{1mm}Gerd Scholl}\\
	Electrical Measurement Engineering\\
	Helmut-Schmidt-University\\
	Hamburg, Germany\\
	\texttt{gerd.scholl@hsu-hh.de} \\
}

\renewcommand{\shorttitle}{Benign Connectivity Variations in Encrypted OPC~UA over Private~5G}

\hypersetup{
	pdftitle={Impact of Benign Connectivity Variations on Intrusion Detection for Encrypted OPC~UA Traffic in Industrial Private~5G Networks},
	pdfsubject={OPC~UA, Intrusion Detection System, Private~5G, Machine Learning},
	pdfauthor={Song Son Ha},
	pdfkeywords={OPC~UA, Intrusion Detection System, Private~5G, Machine Learning},
}

\begin{document}
	\maketitle
	
	
	\begin{abstract}
		\footnote{This is the author's version of a paper that has been accepted for presentation at the 31st IEEE International Conference on Emerging Technologies and Factory Automation (ETFA 2026), to be held in Västerås, Sweden, on September 08–11, 2026.}
		Machine learning (ML)-based intrusion detection systems (IDSs) are increasingly used to monitor encrypted industrial communication. However, their behavior under realistic private~5G operating conditions remains insufficiently understood. This paper investigates the impact of benign connectivity variations on ML-based IDSs for encrypted Open Platform Communications Unified Architecture (OPC UA) traffic in industrial private~5G networks. Experimental results show that legitimate connectivity events can noticeably increase false positive activity despite the absence of attacks. Furthermore, elevated IDS anomaly scores frequently coincide with periods of control-plane (CP) activity associated with these events. The findings highlight the importance of considering CP context when interpreting IDS outputs in industrial private~5G environments.
	\end{abstract}
	
	\acresetall
	
	
	\section{Introduction}
	
	The increasing adoption of private~5G networks in industrial environments enables flexible and reliable connectivity for a wide range of industrial applications. In these environments, OPC~UA is widely used to support industrial communication and commonly employs message-level security mechanisms to protect sensitive industrial data flows \cite{Alwhbi}. While such protection enhances confidentiality, it also limits the effectiveness of traditional IDSs that rely on payload inspection. ML-based IDSs have therefore emerged as promising solutions for OPC~UA intrusion detection. However, these systems are typically trained under baseline conditions, an assumption that rarely holds in private~5G deployments where benign connectivity variations, such as UE reconnection procedures, PDU session resets, and temporary connectivity interruptions, may occur during normal operation. Although these events are not malicious, they can alter traffic patterns and potentially increase false positive rates (FPR). This work investigates the impact of benign connectivity variations on ML-based IDS outputs for encrypted OPC~UA traffic in the absence of attacks. The results show that benign connectivity variations increase false positive activity and that elevated IDS anomaly scores frequently coincide with periods of CP activity. These findings highlight the importance of considering CP context when interpreting IDS outputs in industrial private~5G environments.

	The remainder of this paper is organized as follows. Section~\ref{sec:related_work} reviews related work, Sections~\ref{sec:methodology} and \ref{sec:Implementation} describe the methodology and experimental setup, Section~\ref{sec:results} presents the evaluation results, and Section~\ref{sec:conclusion} concludes the paper and outlines future work.

	\section{Related Work}
	\label{sec:related_work}
	
	Monitoring approaches for encrypted OPC~UA increasingly rely on observable metadata and traffic characteristics rather than decrypted payloads~\cite{honda2022monitoring}. ML-based intrusion detection for encrypted traffic has been widely studied using statistical, timing, and flow-level features as alternatives to deep packet inspection~\cite{ji2024encryptedsurvey}. Recent work further demonstrated effective detection of OPC~UA-specific attacks in industrial private~5G networks using ML techniques~\cite{ha2026}. Related studies have also investigated anomaly detection in 5G infrastructures using network telemetry and CP information~\cite{10436766,wangicpads2023gsad}, while concept drift and contextual factors have been identified as important challenges for reliable IDS operation~\cite{Shyaa2024DriftIDS,stodt2026context}. However, the interaction between benign connectivity variations in private~5G networks, encrypted OPC~UA traffic, and IDS false positive behavior remains insufficiently understood.	To address this gap, this paper analyzes the relationship between CP activity and IDS outputs under benign connectivity variation scenarios.

	\section{Proposed Methodology}
	\label{sec:methodology}
	

	\subsection{Experimental Workflow}
	\label{subsec:work_flow}
	
	All OPC~UA communication is protected using the \texttt{SignAndEncrypt} security mode, preventing payload inspection and requiring intrusion detection to rely on payload-agnostic traffic characteristics. Encrypted OPC~UA traffic observed at the user plane (UP) is segmented into consecutive 5\,s time windows (TWs) and transformed into statistical feature vectors that serve as input to the ML-based IDS. The system is trained using encrypted OPC~UA traffic collected under baseline conditions and subsequently evaluated under controlled benign connectivity variation scenarios to analyze their impact on the FPR. In parallel, UE-level CP signaling traffic is captured and processed to extract lightweight CP indicators. Although not used as IDS inputs, these indicators are aggregated over the same TWs and aligned with IDS outputs to enable a joint analysis of IDS behavior and CP activity under benign connectivity variation scenarios.

	\subsection{Statistical Features from Encrypted OPC~UA Traffic}
	\label{subsec:traffic_features}
	Within each TW, bidirectional OPC~UA flows are identified and processed independently. Flow-level observations are subsequently aggregated across all active flows within the TW to construct a single feature vector representing the overall communication behavior during the observation interval. Statistical features are extracted to characterize transport behavior, temporal dynamics, and OPC~UA lifecycle activity. Transport features describe traffic volume, directionality, asymmetry, packet characteristics, and flow dominance, while temporal features characterize communication rate, timing, duration, variability, activity, entropy, and flow-level dynamics. Observable OPC~UA framing metadata, such as message chunking characteristics and secure channel establishment or closure events, is incorporated to represent lifecycle-related protocol behavior despite payload encryption.	The selected feature categories are adapted from OPC~UA intrusion detection features reported in prior work \cite{ha2026}, retaining only those observable under the \texttt{SignAndEncrypt} security mode.

	\begin{table}[t]
		\centering
		\caption{CP Context Indicators Derived from UE-Level Signaling}
		\label{tab:cp_context}
		\renewcommand{\arraystretch}{1.05}
		\begin{threeparttable}
			\begin{tabular}{
					@{}
					>{\raggedright\arraybackslash}p{0.22\columnwidth}
					@{\hspace{6pt}}
					p{0.75\columnwidth}
					@{}
				}
				\toprule
				\textbf{CP Indicator} & \textbf{Description} \\
				\midrule
				
				Registration Event &
				Binary flag indicating 5G registration or deregistration procedures and changes in UE registration state. \\
				
				\addlinespace[4pt]
				
				CM State Transition &
				Binary flag indicating transitions between CM-IDLE and CM-CONNECTED states. \\
				
				\addlinespace[4pt]
				
				PDU Session State Change &
				Binary flag indicating changes in PDU session availability between active and inactive states. \\
				
				\addlinespace[4pt]
				
				PDU Resource Reconfiguration &
				Binary flag indicating NGAP resource setup, release, or modification procedures for an active PDU session. \\
				
				\addlinespace[4pt]
				
				PDU Session Signaling Count &
				Total number of session-management messages related to PDU session establishment and release. \\
				
				\addlinespace[4pt]
				
				CP Signaling Count &
				Total number of observed CP signaling messages, including NAS and NGAP procedures. \\
				
				\bottomrule
			\end{tabular}
		\end{threeparttable}
	\end{table}

	\subsection{Control-Plane Context Indicators}
	\label{subsec:cp_context}
	
	To provide CP context for the analyzed benign connectivity variation scenarios, a set of lightweight CP indicators is extracted from UE-level signaling traces containing Non-Access Stratum (NAS) and Next Generation Application Protocol (NGAP) messages, as summarized in Table~\ref{tab:cp_context}. The indicators capture connectivity-related activity observable at the CP, including registration procedures, Connection Management (CM) state transitions, Protocol Data Unit (PDU) session events, and overall CP signaling intensity. CP indicator values are aggregated across all observed UEs within each TW and temporally aligned with the corresponding UP traffic features and IDS outputs. This alignment enables a joint analysis of IDS outputs and CP activity under benign connectivity variation scenarios.

	\section{Implementation}
	\label{sec:Implementation}
	
	\subsection{Experimental Environment}
	\label{subsec:hardware}
	
	The experimental evaluation is conducted within the industrial testbed presented in~\cite{11205743}, which incorporates a 3GPP Release~16 compliant private~5G Standalone network, a UP traffic mirroring infrastructure, and an IDS evaluation platform. The OPC~UA traffic generation setup is based on the industrial application environment described in~\cite{ha2026}, with all OPC~UA communication protected using the \texttt{SignAndEncrypt} security mode. The environment includes multiple OPC~UA endpoints representing controller, HMI, and supervisory communication roles, resulting in heterogeneous industrial communication patterns. The collected encrypted traffic is processed using the R\&S®PACE 2 library for payload-agnostic traffic analysis and statistical feature extraction.

	\subsection{Attack Scenarios}
	\label{subsec:attacks}
	
	The evaluation includes four protocol-compliant OPC~UA attack scenarios: Persistent Secure Channel Exhaustion (PSCE), Publish Request Flooding (PRF), Browse Address Space (BAS), and Translate Browse Path (TBP), representing resource exhaustion, excessive service invocation, recursive information model traversal, and complexity-driven service abuse, respectively. For each attack scenario, three operational configurations, denoted as Level~1 (L1), Level~2 (L2), and Level~3 (L3), are defined to represent different attack intensities. L1 corresponds to the highest attack intensity, whereas L3 represents a lower-intensity attack variant that generates communication patterns more similar to normal OPC~UA operation. L2 provides an intermediate operating point between these two extremes. All attack scenarios are implemented by extending Claroty’s OPC~UA exploitation framework~\cite{claroty2025} to support OPC~UA communication protected by the \texttt{SignAndEncrypt} security mode.

	\begin{table}[b]
		\centering
		\caption{Benign Connectivity Variation Scenarios}
		\label{tab:env_variations}
		\renewcommand{\arraystretch}{1.1}
		
		\begin{tabular}{
				@{}
				>{\raggedright\arraybackslash}p{0.23\columnwidth}
				@{\hspace{4pt}}
				p{0.75\columnwidth}
				@{}
			}
			\toprule
			\textbf{Scenario} &
			\textbf{Expected Effects}
			\\
			\midrule
			
			Hard UE Reconnection &
			Registration, security re-establishment, and PDU session recovery with pronounced CP signaling activity.
			\\
			\addlinespace[2pt]
			
			Soft UE Reconnection &
			Deregistration and re-registration signaling with PDU session re-establishment.
			\\
			\addlinespace[2pt]
			
			PDU Session Reset &
			PDU session release and re-establishment with transient UP disruption.
			\\
			\addlinespace[2pt]
			
			PDU Session Interruption &
			Temporary UP interruption without full session release, potentially triggering CM state transitions.
			\\
			
			\bottomrule
		\end{tabular}
	\end{table}

	\subsection{Benign Connectivity Variation Scenarios}
	\label{subsec:env_variation}
	
	To evaluate the impact of benign connectivity variations on encrypted OPC~UA intrusion detection in private~5G networks, a set of connectivity variation scenarios is introduced, including hard and soft UE reconnection procedures as well as PDU session reset and interruption events, as summarized in Table~\ref{tab:env_variations}. These scenarios reflect plausible connectivity events arising from legitimate industrial network operation that may temporarily modify communication patterns and alter encrypted traffic behavior without representing malicious activity, thereby potentially affecting IDS outputs and FPR. To assess intrusion detection robustness under realistic deployments, the connectivity variation scenarios are excluded from model training and used solely during evaluation.

	\subsection{Dataset Construction}
	\label{subsec:dataset_anomaly_detection}
	
	The dataset comprises encrypted OPC~UA traffic collected under baseline conditions, benign connectivity variation scenarios, and malicious attack scenarios. All traffic traces are stored as Packet Capture (PCAP) files with a duration of approximately 10\,minutes. The dataset contains 90 baseline PCAP files, 40 PCAP files collected under benign connectivity variation scenarios, and 120 malicious PCAP files generated from four OPC~UA attack types. Each benign connectivity variation scenario comprises ten PCAP files, with two executions of the event per PCAP. Each attack type is implemented under three operational configurations, with ten PCAP files collected per level. For ML-based IDS development, baseline and malicious traffic are partitioned at the capture level into training, validation, and test sets. Baseline PCAP files are divided into 63 training, 9 validation, and 18 test files. Malicious PCAP files are divided into 60 training, 24 validation, and 36 test files while preserving attack configuration distributions.

	\subsection{Machine Learning Models}
	\label{subsec:ml_models}
	
	Intrusion detection is formulated as a supervised binary classification problem distinguishing benign from malicious encrypted OPC~UA traffic behavior. Four supervised ML models are evaluated: Logistic Regression (LogReg), Random Forests (RF), Support Vector Machines with radial basis function kernels (SVM), and Extreme Gradient Boosting (XGBoost). Hyperparameter tuning is performed using grid search with GroupKFold cross-validation on the training set. The decision threshold is selected on the validation set by maximizing the $F_1$-score while constraining the baseline FPR to 5\%. Final performance is evaluated on the test set. Recall is used to assess intrusion detection performance, while FPR quantifies the impact of benign connectivity variation scenarios.

	\begin{table}[b]
		\centering
		\caption{Recall across attack types and intensity levels.}
		\label{tab:per_level_recall}
		\renewcommand{\arraystretch}{1.1}
		
		\begin{tabular}{llcccc}
			\toprule
			\textbf{Attack Type} &
			\textbf{Level} &
			\textbf{LogReg} &
			\textbf{RF} &
			\textbf{SVM} &
			\textbf{XGBoost} \\
			\midrule
			
			\multirow{3}{*}{BAS}
			& L1 & 0.892 & 0.899 & \textbf{0.912} & 0.899 \\
			& L2 & 0.506 & 0.637 & 0.630 & \textbf{0.659} \\
			& L3 & 0.349 & 0.430 & 0.442 & \textbf{0.454} \\
			\midrule
			
			\multirow{3}{*}{PSCE}
			& L1 & 0.854 & 0.847 & 0.835 & \textbf{0.856} \\
			& L2 & 0.659 & 0.680 & \textbf{0.697} & 0.692 \\
			& L3 & 0.248 & 0.311 & 0.316 & \textbf{0.321} \\
			\midrule
			
			\multirow{3}{*}{PRF}
			& L1 & 0.812 & 0.823 & 0.819 & \textbf{0.833} \\
			& L2 & 0.773 & 0.786 & \textbf{0.817} & 0.808 \\
			& L3 & 0.353 & 0.650 & 0.585 & \textbf{0.674} \\
			\midrule
			
			\multirow{3}{*}{TBP}
			& L1 & 0.824 & 0.829 & \textbf{0.834} & 0.824 \\
			& L2 & 0.496 & 0.644 & 0.658 & \textbf{0.663} \\
			& L3 & 0.324 & 0.384 & \textbf{0.394} & 0.384 \\
			
			\bottomrule
		\end{tabular}
	\end{table}

	\section{Evaluation Results} 
	\label{sec:results}
	
	\subsection{Detection Performance}
	
	Table~\ref{tab:per_level_recall} summarizes the recall achieved by the evaluated ML-based IDS models for the considered OPC~UA attack scenarios and operational configurations under encrypted OPC~UA communication. All evaluated models achieve high recall for the highest-intensity attack configurations, demonstrating that malicious activity remains detectable even when OPC~UA payloads are protected using the \texttt{SignAndEncrypt} security mode. Recall decreases progressively as attack intensity is reduced from L1 to L3. This trend reflects the attack design, as lower-intensity configurations produce less pronounced changes in the encrypted OPC~UA traffic characteristics, resulting in a more challenging classification task. The decrease in recall is particularly evident for BAS, PSCE, and TBP, whose L3 recall values remain below 0.46 even for the best-performing model. In contrast, PRF remains comparatively easier to detect, with XGBoost achieving a recall of 0.674 for the L3 configuration.

	\begin{figure}[t]
		\centering
		\includegraphics[width=0.68\linewidth]{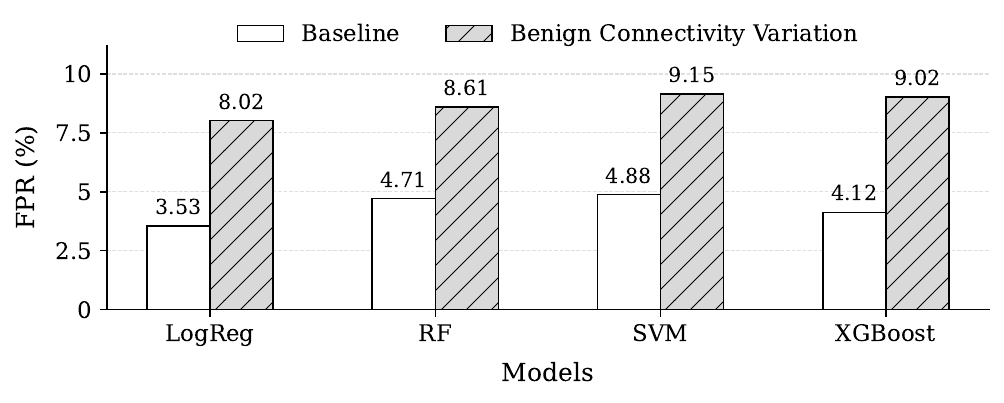}
		\caption{Impact of Benign Connectivity Variations on FPR.}
		\label{fig:FPR}
	\end{figure}
	
	\begin{figure}[t]
		\centering
		\includegraphics[width=0.68\linewidth]{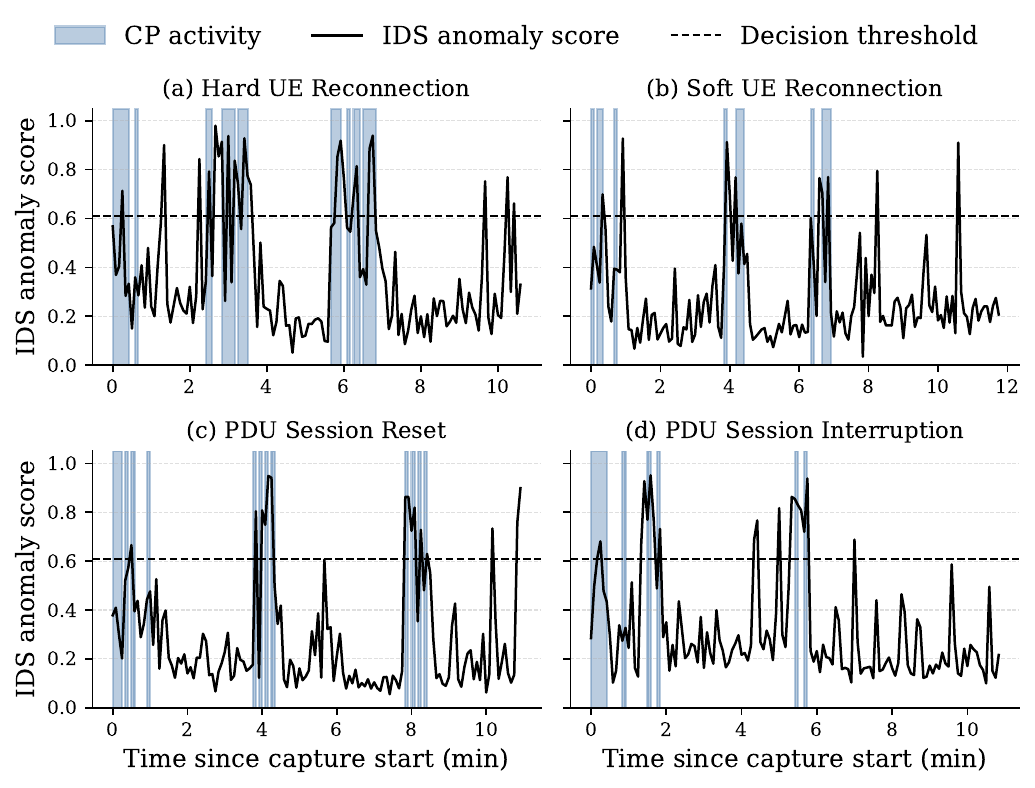}
		\caption{Temporal relationship between CP activity and IDS anomaly scores under benign connectivity variation scenarios.}
		\label{fig:Alignment}
	\end{figure}

	\subsection{Impact of Benign Connectivity Variations}
	
	To evaluate the impact of benign connectivity variations on false positive activity, the trained models are evaluated on both the baseline test subset and the benign connectivity variation dataset. The resulting FPRs are shown in Fig.~\ref{fig:FPR}. Under benign connectivity variation scenarios, all evaluated ML-based IDS models exhibit higher FPR values, increasing from 3.53\%--4.88\% under baseline conditions to 8.02\%--9.15\%. These results indicate that benign connectivity variations can substantially influence IDS outputs despite the absence of malicious traffic. Furthermore, the comparable FPR increase observed across all four evaluated model families suggests that the phenomenon is largely systematic rather than specific to a particular classifier architecture or learning paradigm.

	\subsection{Control-Plane Events and False Positive Activity}
	
	To investigate the relationship between benign connectivity events and IDS outputs, CP-active TWs are compared with IDS anomaly scores. As XGBoost achieves the highest overall recall in Table~\ref{tab:per_level_recall}, it is selected for the subsequent analysis. Fig.~\ref{fig:Alignment} illustrates the temporal relationship between CP-active TWs and IDS anomaly scores using four representative recordings, one from each benign connectivity variation scenario summarized in Table~\ref{tab:env_variations}. A CP-active TW is defined as a 5\,s TW in which at least one of the CP indicators listed in Table~\ref{tab:cp_context} is observed. Across all four scenarios, CP-active TWs are concentrated within three distinct phases. The first phase occurs near the beginning of each recording and is associated with normal network operation and UE activity. The remaining two phases correspond to the two controlled executions of the corresponding connectivity variation scenario. 
	
	For all four scenarios, IDS anomaly scores frequently increase within or shortly after the identified CP-active phases. In many cases, the anomaly score exceeds the decision threshold during these phases, indicating a temporal association between false positive detections and benign connectivity events. This behavior is observed across all evaluated benign connectivity variation scenarios. At the same time, occasional threshold crossings can also be observed outside the identified CP-active phases. This observation is consistent with the non-zero baseline FPR reported in Fig.~\ref{fig:FPR}, indicating that not all false positive detections are associated with CP activity.

	\section{Conclusion and Future Work}
	\label{sec:conclusion}
	
	This paper investigated the impact of benign connectivity variations on ML-based IDS for encrypted OPC~UA traffic in private~5G networks. Experimental results showed that the evaluated ML models maintain reasonable detection performance under encrypted communication. However, benign connectivity variations substantially increased the FPR across all evaluated models despite the absence of malicious traffic. Furthermore, temporal analysis revealed that elevated IDS anomaly scores often occur during or shortly after CP-active TWs associated with benign connectivity events. These findings suggest that benign connectivity variations should be considered when designing and deploying ML-based IDSs in industrial private~5G environments.
	
	Future work will investigate CP-aware intrusion detection approaches that combine encrypted traffic features with CP context information to improve robustness against benign connectivity variations and reduce false positive activity.

	\section*{Acknowledgment}
	
	The authors would like to thank K. Singh, J. Jockram, F. Mueller, ipoque GmbH, Deutsche Telekom, and Ericsson for their continuous support and valuable cooperation throughout this work.
	
	\section*{Funding}
	This research is funded by dtec.bw – Digitalization and Technology Research Center of the Bundeswehr. dtec.bw is funded by the European Union – NextGenerationEU (project “Digital Sensor-2-Cloud Campus Platform” (DS2CCP), https://dtecbw.de/home/forschung/hsu/projekt-ds2ccp).
	
	\bibliographystyle{unsrtnat}
	\bibliography{references}
	
\end{document}